# ATMOSPHERIC MASS LOSS FROM HOT JUPITERS IRRADIATED BY STELLAR SUPERFLARES

**D. V. Bisikalo[1], V.I. Shematovich[1], A.A. Cherenkov[1], L. Fossati[2], C. Möstl[2]**

[1]*Institute of Astronomy, Russian Academy of Sciences, Moscow, Russia*

[2]*Space Research Institute, Austrian Academy of Sciences, Graz, Austria*

**Abstract.** Because of their activity, late-type stars are known to host powerful flares producing intense high-energy radiation on short time-scales that may significantly affect the atmosphere of nearby planets. We employ a one-dimensional aeronomic model to study the reaction of the upper atmosphere of the hot Jupiter HD 209458b to the additional high-energy irradiation caused by a stellar flare. Atmospheric absorption of the additional energy produced during a flare leads to local atmospheric heating, accompanied by the formation of two propagating shock waves. We present estimates of the additional atmospheric loss occurring in response to the flare. We find the mass loss rate at the exobase level to significantly increase ($3.8 \times 10^{10}$, $8 \times 10^{10}$, and $3.5 \times 10^{11}$ g s$^{-1}$ for 10, 100, and 1000 times the high-energy flux of the quiet star, respectively) in comparison to that found considering the inactive star ($2 \times 10^{10}$ g s$^{-1}$).

*Key words*: stars: flare - planet–star interactions - planets and satellites: atmospheres - planets and satellites: physical evolution

## Introduction

The activity characterizing late-type stars gives rise to flares and coronal mass ejections that are believed to have an important influence on the evolution of the atmosphere of planets, particularly if close-in (e.g., Luger et al. 2015; Cherenkov et al. 2017; Tilley et al., 2018). Because stellar activity increases with decreasing stellar age, the effects of these extreme phenomena on the atmosphere of planets are particularly important for young systems.

Analyses conducted on the basis of Kepler 1-minute cadence observations (Maehara et al. 2012, 2015; Shibayama et al. 2013) revealed that flares are found ubiquitously among late-type stars. In addition, ultraviolet observations have also shown that the number and intensity of flares is particularly high for M dwarfs (France et al. 2016). The Kepler data allowed also the detection of 187 superflares on 23 solar-type stars whose bolometric energy ranges between $10^{32}$ and $10^{36}$ erg (Maehara et al. 2015). Some superflares present multiple peaks separated by 100 to 1000 seconds, which is comparable to the period of quasi-periodic pulsations in solar and stellar flares (Maehara et al. 2015; Davenport 2016). The data allowed to conclude that the frequency (dN/dE) of superflares as a function of flare energy (E) could be described by a power-law distribution dN/dE ~ $E^{-\alpha}$ with $\alpha$ ~ $-1.5$ for $10^{33} < E < 10^{36}$ erg flares (Candelaresi et al. 2014; Maehara et al. 2015). The average occurrence rate of superflares with an energy in excess of $10^{33}$ erg, which is equivalent to 100 times a typical solar flare, is about once in 500-600 years (Maehara et al. 2015). It has also been shown that the duration of superflares ($\tau$) increases with flare energy as $\tau \sim E^{0.39 \pm 0.03}$ (Maehara et al. 2015).

Multi-wavelength observations on solar proxies with ages between 100 Myr and 7 Gyr show that the young Sun was likely rotating more than 10 times its present rate and therefore possessed a strong dynamo-driven high energy emission (Ribas et al. 2003). In particular, the results of the "Sun-in-Time" project suggest that the coronal hard X-ray (0.1–2 nm), soft X-ray (SXR, 2 - 10 nm) and extreme ultraviolet (EUV, 10-91.2 nm) emission of young solar twins were respectively about 1550, 130, 80 times stronger than those of the present Sun (e.g., Lammer et al. 2012). This increased activity level leads also to a higher rate and energy of stellar flares and coronal mass ejections.

For giant planets, the high-energy stellar radiation (X-ray+EUV, hereafter XUV) is mostly absorbed in the thermosphere, which usually lies at altitudes ranging between 1.1 and 1.5 planetary radii ($R_{pl}$; Yelle 2004; Shematovich et al. 2014). Bisikalo et al. (2018) employed a recently developed one-dimensional (1D) aeronomic model of the upper atmosphere of hot Jupiters and Neptunes (Ionov et al. 2017, 2018) to study the effects of an average flare on the atmosphere of the hot Jupiter HD209458b. They showed that the atmospheric absorption of the extra-energy provided by a stellar flare leads to a substantial local heating of the planetary atmosphere followed by the formation of two shock waves propagating along the atmosphere. They further presented the increase in planetary Ly-α luminosity caused by the impact and absorption of the energy emitted by a stellar flare.

We present here estimates of the atmospheric mass loss rates of the hot Jupiter HD209458b subject to different kinds of stellar flares. This paper is organized as follows. Section 2 presents the input parameters of the model and the reaction of the atmospheric structure to the stellar flares. In Section 3, we derive the induced mass loss rates, while in Section 4 we discuss the importance of the results in the context of planetary evolution and gather our conclusions.

## 2. Planetary atmosphere modeling

We follow the approach of Bisikalo et al. (2018) to simulate the response to superflares of the atmosphere of the hot Jupiter HD209458b. We model the flare as a short-term increase in the intensity of the stellar XUV radiation. We consider three different types of flares with different intensities and durations. The weakest flare lasts for 12 minutes and it corresponds to an increase of 10 times the non-flaring solar XUV flux with energy flux of 2.7 erg cm$^{-2}$s$^{-1}$ at 1 a.u. For all calculations, the stellar XUV flux is scaled to the planetary orbital distance. We further consider two stronger flares with intensities of 100 and 1000 times that of the quiet star, with durations of 24 and 30 minutes, respectively (Maehara et al. 2015). We assume that during a flare the intensity of the radiation increases equally across the whole XUV wavelength range. We further assume that the flare remains at its given intensity for its whole duration, thus without increasing at the beginning and gradually decreasing towards the end. Because of the short flare duration, this assumption does not affect the final results.

We model the planetary atmosphere employing the code of Ionov et al. (2017, 2018). This is a self-consistent 1D code taking into account both atmospheric heating caused by atmospheric absorption of stellar XUV radiation and the contribution of reactions involving suprathermal photoelectrons. The code includes three main modules: a Monte Carlo module, a module of chemical kinetics and a gas-dynamic one. In the Monte Carlo module, the heating rate of the atmosphere, rates of photolytic reactions, and the rates of reactions caused by suprathermal photoelectrons are computed by solving the Boltzmann equation for photo- and secondary electrons (Shematovich et al. 2014). Using rates of photolytic reactions obtained by the Monte-Carlo module, the chemical kinetics module solves the system of equations of chemical kinetics and computes the number density of the atmospheric components in every cell. The gas-dynamic module computes the profiles of the macroscopic atmospheric parameters, such as density, velocity, and temperature. The gravitational potential was set equal to the three-dimensional Roche potential along the line connecting the planetary and the stellar centers.

The use of a 1-D model has several limitations, the most important being neglecting atmospheric circulation. Accounting for this effect can modify the atmospheric profiles and influence the mass-loss rate. A further limitation of our approach is that we do not account for a planetary magnetic field. Recent studies (e.g., Arakcheev et al., 2017; Bisikalo et al., 2017) have shown that planetary magnetic fields significantly influence the gas dynamics in the upper atmosphere, where the gas is highly ionized. In this paper we consider mostly the lower atmospheric layers (below the exobase) where the ionization rate is not too high, thus the assumption of a non-magnetised planet does not have a strong influence on the results.

Figures 1, 2, and 3 show the atmospheric temperature (top), mass density (middle), and bulk velocity (bottom) profiles obtained for the three considered flares at six different moments in time from the beginning of the flare. The exobase for the adopted parameters is located at about 3.5 $R_{pl}$, while the Knudsen number approaches a value of about 1 at roughly 4.5$R_{pl}$. This justifies that the atmospheric profiles have been computed (and shown) up to 4.5$R_{pl}$. The behavior of the gas in the three different runs is qualitatively the same, we therefore describe in detail only the one obtained with the flare having an XUV flux 100 times that of the quiet star (Figure 1). The atmospheric temperature profile for the steady-state solution presents a maximum of about 7000 K at about 1.3 $R_{pl}$ (black line). The heating caused by the absorption of the XUV flux emitted by the quiet star leads to a slow (~2.5 km/s) hydrodynamic expansion of the atmosphere, as shown by the increase of the bulk velocity above 1.3 $R_{pl}$. Our model provides a slightly lower thermosphric temperature comparing with the previous aeronomic models (Yelle 2004; Koskinen et al. 2013). Details of such comparison can be found in (Ionov et al., 2017). Once the flare reaches the planet, the temperature maximum increases to 9000 K and the position of the peak slightly shifts to a higher altitude relative to the initial position. As a consequence of the heating, these layers begin to expand and most of the absorbed energy of the flare is transformed into kinetic energy driving the expansion. The velocity of the expanding atmosphere reaches about 15 km/s and as soon as it becomes supersonic a shock wave forms, which accelerates outwards with an average speed of about 23 km/s. After 4 hrs, the shock reaches the upper boundary at about 4.5$R_{pl}$. As the first shock moves away from the lower thermosphere, the matter with an excess kinetic energy moves upwards, falling then down once cooled. The in-falling material becomes supersonic about 2 hrs after the flare, leading to the formation of a second shock wave. This wave is quite week and slow (it travels outwards with a velocity of about 5 km/s). Once this second shock wave has disappeared, the atmosphere returns to the initial condition. The atmosphere returns to the initial condition about 4 hrs after the flare hit.

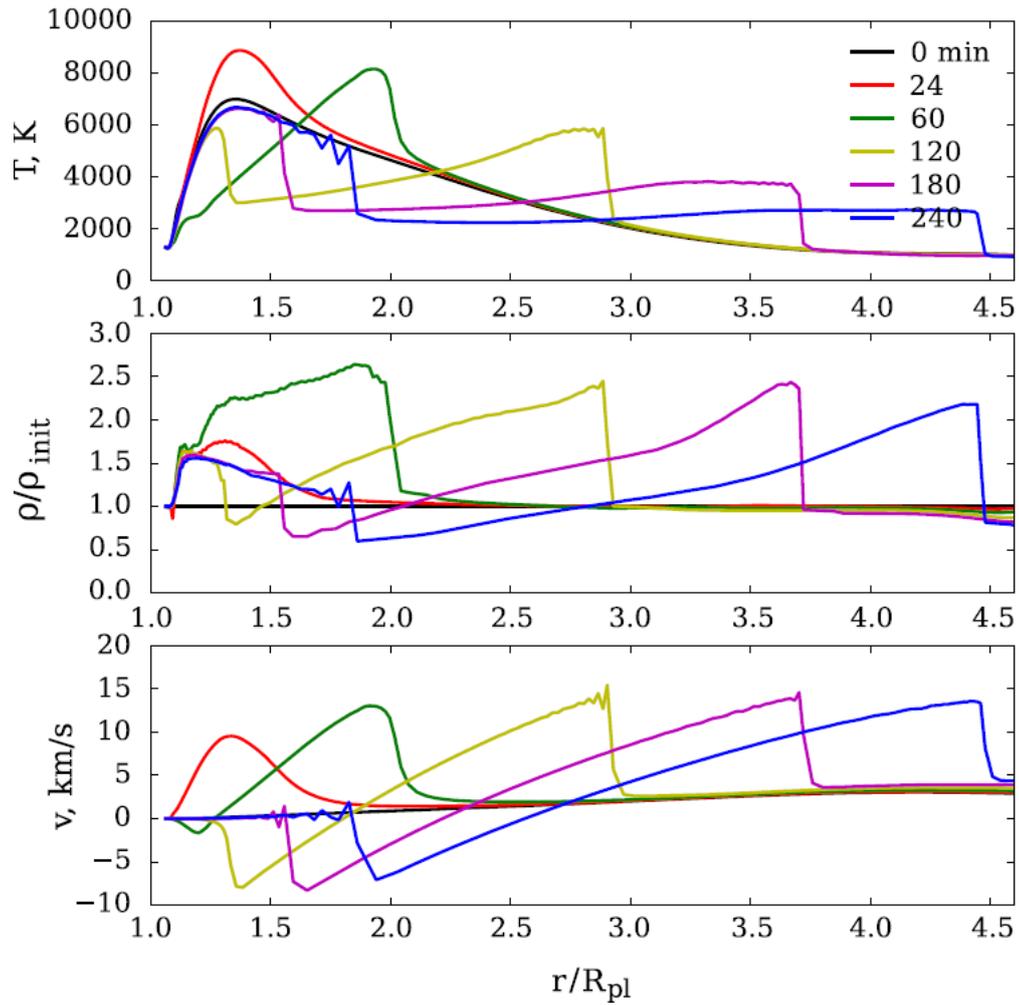

Figure 1. Radial profiles for temperature (top), mass density relative to the initial state (middle) and bulk velocity (bottom) in the atmosphere of the hot Jupiter HD 209458b during the hit of a stellar flare with an XUV flux 100 times that of the quiet star. Each line (see legend) corresponds to a different time following the flare hit.

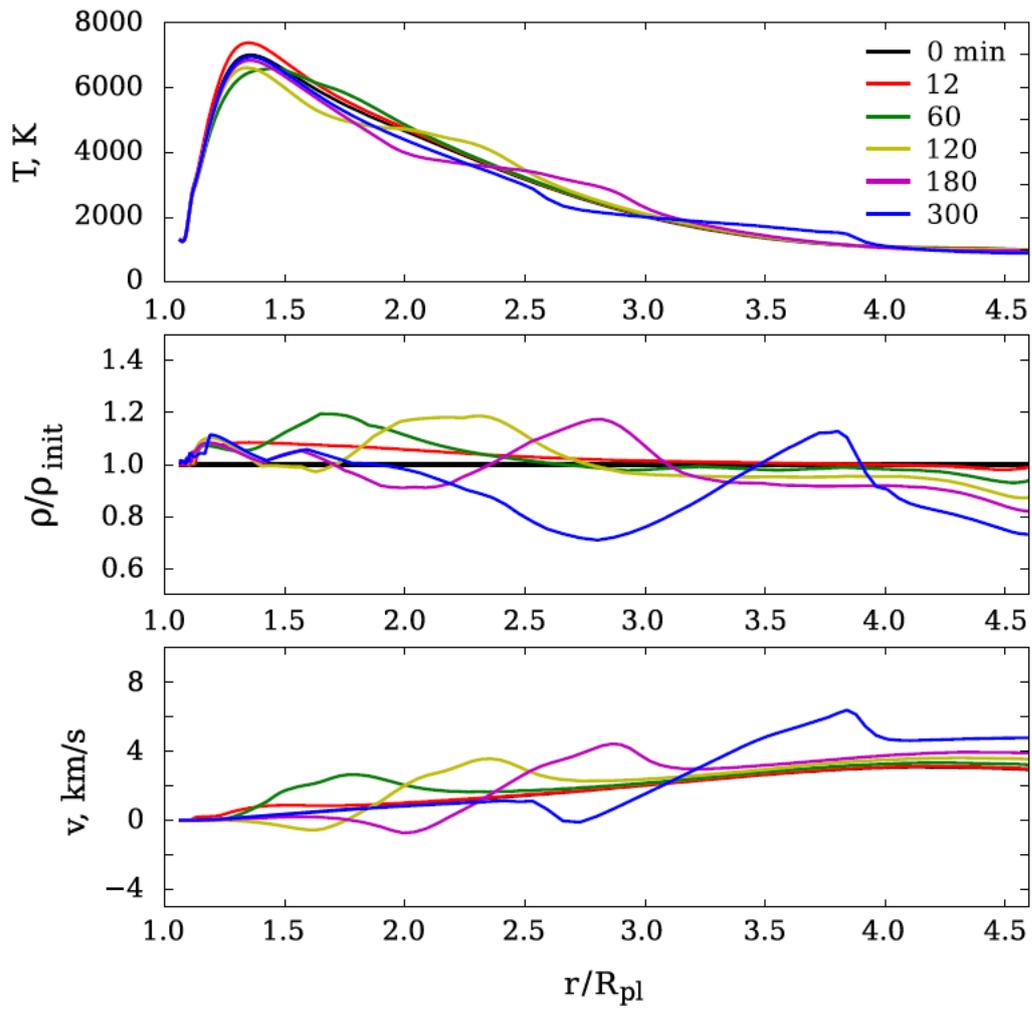

Figure 2. Same as Figure 1, but for a flare with an XUV flux 10 times that of the quiet star.

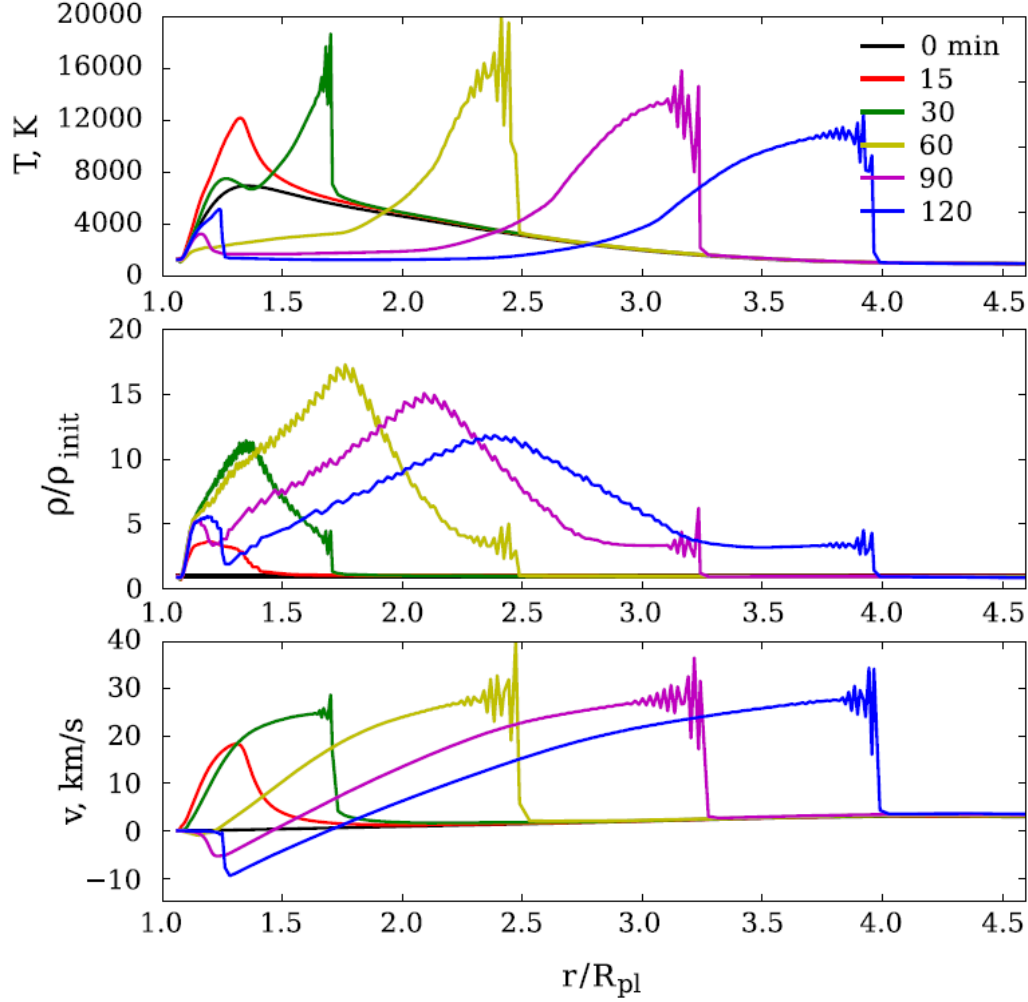

Figure 3. Same as Figure 1, but for a flare with an XUV flux 1000 times that of the quiet star.

We obtain similar results also for the other two considered flares (Figures 2 and 3), except for the case of the weaker flare for which the second shock wave does not form due to the fact that the in-falling atmospheric gas does not reach a high enough velocity.

## 3. Mass-loss rates induced by superflares.

The development of shock waves significantly affects atmospheric loss. We present this in Fig. 4, which shows the mass-loss rates at the exobase as a function of time for all three considered superflares. For the case of the quiet star, we find a mass-loss rate of $2\times10^{10}$ g s$^{-1}$ in the spherical symmetry approach, which increases by 1.9, 4, and 17.5 times during a flare emitting an XUV flux that is 10, 100, and 1000 times that of the quiet star, respectively. Since the excess mass loss occurs for 130, 210, and 380 minutes, the additional atmospheric loss corresponds to $3\times10^{14}$, $1\times10^{15}$, and $8\times10^{15}$ g for the three considered flares, respectively.

Depending on the frequencies of the flares the total mass loss from the atmosphere might therefore significantly increase. For old solar-type stars, taking into account that flares emitting an XUV flux 10, 100, and 1000 times that of the quiet star are believed to happen approximately 1, 0.1, and 0.001 times per year, the effect of flares on the evolution of planetary atmospheres can be considered to be negligible. However, the impact of flares may be more pronounced for younger systems.

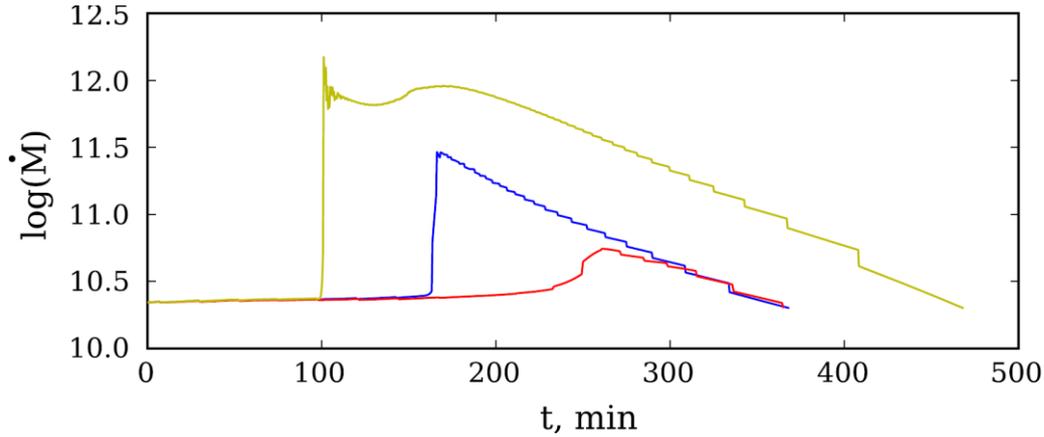

Figure 4. Time evolution of the mass-loss rates (in g s$^{-1}$) obtained at the exobase for the atmosphere of a hot Jupiter like HD 209458b during the three considered superflares. The red, blue, and green lines correspond to the mass-loss rates obtained considering flares emitting an XUV flux of 10, 100, and 1000 times that of the quiet star, respectively.

## 4. Conclusions.

Results of the simulations show that additional local **heating** caused by a flare leads to the formation of two propagating outward shock waves. The first fast wave lead to the expansion of the atmosphere and to the additional atmospheric loss occurring in response to the flare. The mass-loss rate increases by 1.9, 4, and 17.5 times during a flare emitting an XUV flux that is 10, 100, and 1000 times that of the quiet star, respectively, in comparison with case of quiet star of $2\times10^{10}$ g s$^{-1}$.

Taking into account the duration of the mass loss excess of 130, 210, and 380 minutes for the three considered flares, respectively, we found that the additional atmospheric loss corresponds to $3\times10^{14}$, $1\times10^{15}$, and $8\times10^{15}$ g. It means, that for old solar-type stars the effect of flares on the evolution of planetary atmospheres is not important, while for younger systems it could be more pronounced.

The local effect of the flare can be significant and change the gas dynamic structure of the whole envelope. In order to consider the real dynamics of the matter expanding from the atmosphere and its interaction with the stellar wind, it is necessary to use a coupled three-dimensional gas dynamic model, which in turn will allow us to correctly calculate the increase of the atmospheric loss rate from the gaseous envelope of hot Jupiter under the forcing by the stellar superflare. We work on this for the case of semi-open atmospheres (Bisikalo et al, 2013) and will present the results soon.

**Acknowledgements**

This study was supported by the Russian Science Foundation (project No. 18-12-00447). C. Mőstl thanks the Austrian Science Fund (FWF): [P26174-N27] for support.